\documentclass[letterpaper, 10 pt, conference]{ieeeconf}  

\IEEEoverridecommandlockouts                              

\usepackage[
    style=ieee,
    doi=false,
    isbn=false,
    url=false,
    eprint=false,
    backend=bibtex,
    natbib=true
    ]{biblatex}
\usepackage{csquotes}
\usepackage{graphicx}
\usepackage{subcaption}

\usepackage{tikz}
\usetikzlibrary{shapes,arrows,calc,positioning,fit}

\renewcommand{\cite}{\parencite}

\usepackage{amsmath}
\usepackage{amsfonts}

\usepackage{perl_acronyms}
\usepackage{perl_misc}
\newcommand{\subparagraph}{}
\let\labelindent\relax
\usepackage{filecontents}
\usepackage{silence}
\usepackage{siunitx}
\usepackage[inline]{enumitem}

\title{\LARGE \bf
Parametric Design of Underwater Optical Systems
}

\author{Gideon Billings$^{1}$, Eduardo Iscar$^{1}$, Matthew Johnson-Roberson$^{1}$
\thanks{ $^{1}$E. Iscar, G. Billings  and M. Johnson-Roberson are with the Department of Naval Architecture and Marine Engineering, University of Michigan, Ann Arbor, MI 48109 USA
{\tt\small \{eiscar,gidobot, mattjr\}@umich.edu}}
}
\date{June 2016}

\begin{document}
\pagenumbering{gobble}
\setlength{\parskip}{0pt plus 0.0pt minus 2.0pt}
\setlength{\floatsep}{5pt plus 1.0pt minus 0pt}
\setlength{\textfloatsep}{5pt plus 1.0pt minus 5pt}
\addtolength{\topmargin}{-10pt}

\maketitle

\begin{abstract}
The design of optical systems for underwater vehicles is a complex process where the selection of cameras, lenses, housings, and operational parameters greatly influence the performance of the complete system. Determining the correct combination of components and parameters for a given set of operational requirements is currently a process based on trial and error as well as the specialized knowledge and experience of the designer. In this paper, we introduce an open-source tool for the parametric exploration of the design space of underwater optical systems and review the most significant underwater light effects with the corresponding models to estimate the response and performance of the complete imaging system.

\end{abstract}

\section{Introduction}
\label{sec:motivation}
Optical cameras are increasingly being applied in the underwater domain for a range of applications including inspection tasks~\cite{calvo2008experimental}, ecosystem monitoring~\cite{williams2012monitoring} and vehicle navigation~\cite{eustice2008visually}.  Cameras represent low cost, low power sensors that provide rich information about the underwater scene and frequently complement other sensors deployed on \acp{AUV} or \acp{ROV}. However, the design of an underwater camera system presents a very large space of possible design choices and system configurations, with many inter-dependencies. Additionally, field tuning of the camera settings is frequently cumbersome and time consuming due to reduced equipment accessibility when deploying underwater.  

In this paper we review a simplified underwater image formation model that allows the estimation of the average camera sensor response given different lens, light, water and seafloor characteristics. The sensor response is the average intensity of pixels in a camera image and is a metric that can be used to determine correct image exposure. A user-friendly interface for the model is developed that will allow researchers and scientists to narrow down the equipment requirements and operational settings for an underwater imaging system by parametrically exploring the design space.

In order to estimate the camera response, a model of underwater image formation is required. One of the main drivers for the study of the underwater image formation process and the development of models has been the need to correct underwater image degradation such as haze, low contrast and color cast due to water impurities and wavelength dependent attenuation. Early efforts by Duntley~\cite{duntley_light_1963} laid the foundation for modelling underwater light propagation. Computer models developed by  McGlamery~\cite{mcglamery1975computer, mcglamery1980computer} were extended by Jaffe in 1990~\cite{jaffe1990computer}, leveraging advances in computational processing capabilities to create the UNCLES computer simulation system, which is capable of analyzing the performance of underwater camera systems. The UNCLES simulator helped guide the design of the video equipment for the ARGO underwater imaging platform~\cite{jaffe2016sea}, but the tool was not released for public use. The theory for underwater light propagation has previously been developed, but there lacks a consolidation of this knowledge into a framework broadly usable by the science and engineering communities for the design of underwater camera systems. The tool introduced in this paper incorporates the model developed through these prior works with an interface focused on user friendliness and minimal complexity. Some assumptions are made to simplify the model, based on common characteristics of underwater imaging systems, and the validity of this model is demonstrated through experimentation. The contributions presented in this work are
\begin{enumerate*}
    \item A review of the underwater image formation model with a procedure to characterize underwater camera systems.
    \item An open source tool to aid the design process for an underwater camera system through exploration of the parameter space.
    \item Validation experiments supporting the presented model as a good characterization of an underwater camera system.
\end{enumerate*}

The rest of the paper is structured as follows:  Section~\ref{sec:methodology} introduces the underwater image formation model used in the software toolbox to compute sensor responses underwater; Section~\ref{sec:software} presents the developed software toolbox, with an overview of the intended design use and user interface; Section~\ref{sec:experiments} presents the experiments validating the proposed image formation model; and Section~\ref{sec:conclusion} outlines our conclusions and future work.

\begin{figure}[]
    \centering
    \includegraphics[width=0.6\columnwidth]{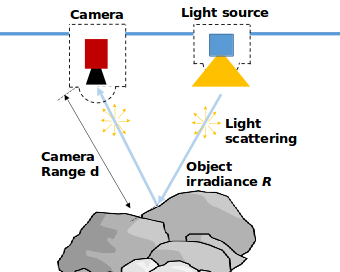}
    \caption{Schematic of underwater light propagation from light source to camera sensor, where the light signal is affected by scattering and absorption through the water column and the reflection characteristics of the seafloor.}
    \label{fig:underwaterlightprop}
\end{figure}

\section{Underwater Image Formation}
\label{sec:methodology}

\begin{figure}[ht!]
\resizebox{\linewidth}{!}{ \tikzstyle{block} = [rectangle, draw, fill=blue!20, 
      text centered, rounded corners, node distance=0.5cm, every node/.style={scale=0.15}, text width=70mm]
 \tikzstyle{arrow} = [thick,->,>=stealth]

 \begin{tikzpicture}[]
     \node[block] (light) {Light Source};
     \node [block,  below=of light] (water_1)  {Underwater Attenuation (Eq.~\ref{eq:uwattenuation})};
     \node [block,  below=of water_1] (object)  {Scene Reflectance (Eq.~\ref{eq:brdf})};
     \node [block,  below=of object] (water_2)  {Underwater Attenuation (Eq.~\ref{eq:uwattenuation})};
     \node [block, below=of water_2] (lens) {Lens (Eq.~\ref{eq:lensradiometric})};
     \node [block,   below=of lens] (sensor) {Image Sensor (Eq.~\ref{eq:absorbedphotons})};
     \node [block,   below=of sensor] (digitization) {Digitization (Eq.~\ref{eq:adcconversion})};
     \node [block,   below=of digitization] (image) {Digital Image };

     \draw [arrow] (light) -- node[anchor= south] {} (water_1);
     \draw [arrow] (water_1) -- node[anchor= south] {} (object);
     \draw [arrow] (object) -- node[anchor= south] {} (water_2);
     \draw [arrow] (water_2) -- node[anchor= south] {} (lens);
     \draw [arrow] (lens) -- node[anchor= south] {} (sensor);
     \draw [arrow] (sensor) -- node[anchor= south] {} (digitization);
     \draw [arrow] (digitization) -- node[anchor= south] {} (image);

\end{tikzpicture}}
\caption{Image formation pipeline describing the different steps through which light is subjected to form the underwater digital image.}
\label{fig:flowchart}
\end{figure}
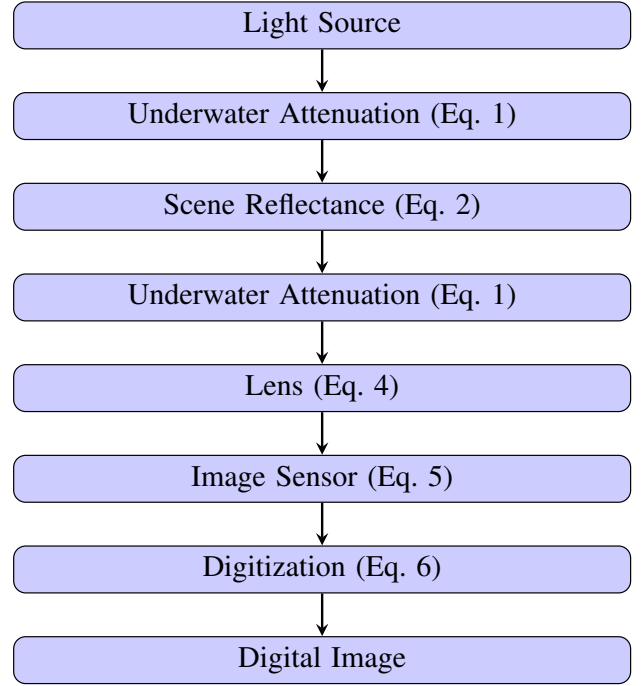

In this section we introduce the underwater image formation model. As light travels from a source through the water column, it is attenuated through absorption and scattering. The light that reaches the seafloor or other obstacle is reflected by a fractional amount, dependant on the albedo of the surface. The reflected light is further attenuated in the water column as it travels back towards the camera. Light is refracted at the water interface of the camera housing viewport before reaching the camera lens. Photons passing through the lens generate electrical signals on the camera sensor that are amplified and digitized to form the final image. This process is illustrated in  Figure~\ref{fig:underwaterlightprop}, and Figure~\ref{fig:flowchart} provides an overview of how the model equations describe the image formation pipeline.

\subsection{Artificial Light systems}
Natural light is attenuated exponentially in the oceans and frequently does not penetrate deeper than $100$m. Our model assumes all light in the scene is generated from artificial light sources mounted on the vehicle. This situation represents the worst case scenario, as constraints on camera systems are relaxed if natural light is present. The presented model describes a light source by three main parameters: 
\begin{enumerate}[wide, labelwidth=!, labelindent=0pt]
    \item Luminous flux emitted by the light source, measured in lumens: This can be obtained for most underwater lights, strobes or LED modules in custom designs. 
    \item Normalized light spectrum: The spectrum of the light source describes how the luminous flux is spread over the different wavelengths. When the spectrum is not available, it can be approximated based on known spectra for common light sources. Figure~\ref{fig:lightsspectrum} shows  spectrum characteristics of common light types such as LED, fluorescent or natural sunlight. 
    \item Beam pattern: The beam pattern describes how the light spreads as it travels away from the source. We assume a simple conical beam pattern defined by its aperture half-angle $\beta$, which is typical for most underwater strobes. 
\end{enumerate}

\subsection{Underwater Light Propagation}
Light traveling underwater from the strobe to the camera sensor is modified through absorption, scattering, reflection, and refraction at optical interfaces. We describe how each of these effects is modeled in our system. 

\subsubsection{Attenuation}
The Jaffe-McGlamery model describes the propagation of light underwater as the sum of direct, backscatterred and forward-scatterred light. Attenuation of the light signal is modeled as an exponential decay, with function parameters depending on the water type and clarity. Coefficients describing the attenuation effects for different classes of water, known as Jerlov water bodies, have been cataloged~\cite{Solonenko:15}. The exponential decay modeling attenuation of the light signal in water is given as
\begin{equation}
    L=Re^{-b(\lambda)d}
    \label{eq:uwattenuation}
\end{equation}
where R is the initial irradiance, $b(\lambda)$ the wavelength dependent attenuation coefficient and d the distance of propagation. Absorption and scattering coefficients are mostly dependent on chlorophyll and dissolved organic matter in the water column~\cite{Solonenko:15}. Experiments performed by Jerlov~\cite{jerlov1951photographic} established a set of attenuation profiles for different types of water bodies, both coastal and oceanic, with varying clarity levels. These profiles are provided with our model as default selections. The user also has the option to load custom profiles.

\subsubsection{Object reflectivity}
The reflectance of light by a surface is modeled  by the \ac{BRDF}~\cite{Nicodemus:65} that relates the outgoing radiance $L$ of the surface with the incoming irradiance $E$. Assuming diffuse reflection in our model, where $\theta_i$ is the light incident angle and $M(\lambda)$ is the material and wavelength dependent reflection coefficient, the \ac{BRDF} is simplified to: 
\vspace{-1mm}
\begin{equation}
L = E\frac{M(\lambda)}{\pi} cos(\theta_i)
\label{eq:brdf}
\end{equation}

\subsubsection{Light refraction}
Underwater cameras are housed inside enclosures that protect the electronic systems from water damage and pressure. In order for light to reach the sensor, these enclosures employ an optical port made of translucent material such as glass or acrylic, most frequently in either a spherical or flat geometry. As light travels through the port, it is refracted at each optical interface as a function of the change in index of refraction and the direction of the incident ray relative to the surface normal. In effect, the optical port of the housing must be considered as part of the camera lens system.

In the case of a domed viewport, the dome is treated as a thick lens formed by two concentric hemispherical surfaces. Analysis of the thick lens equations show that objects at infinity are mapped to a virtual image in the front of the dome that is curved concentrically with the dome~\cite{optics, domeport}. A camera housed with a dome viewport must be focused at the distance of the virtual image when immersed in water rather than the distance to the imaging target in air. The distance of the virtual image from the front of the dome is derived in~\cite{optics, domeport}, and we incorporate these equations into the camera system design tool.

When the camera lens principal point is aligned with the dome center of curvature, the field of view of the camera remains unchanged~\cite{optics, domeport}. A common method to verify the camera is correctly aligned with the center of the dome is to look at an image of a checkerboard taken with the camera in the housing while only half immersed in water. There should be no magnification difference between the part of the image below the water and the part above the water if the camera is centered.

 For the case of flat viewports, the effects of refraction result in a change in the effective lens focal length~\cite{lavest2000underwater}, given as
\begin{equation}
\textit{f}_{uw}=1.33\textit{f}_{air}
\label{eq:effectivefocallength}
\end{equation}
where $f_{uw}$ is the effective focal length in water and $f_{air}$ the focal length in air.
This  increase in the effective focal length of the system reduces the camera field of view and must be accounted for when computing the lens aperture number.


\subsection{Lensing effects}
The fundamental radiometric relation expresses the amount of light incident on the lens that reaches a pixel at the sensor surface~\cite{szeliski2010computer}: 
\begin{equation}
    E_{I} = L\frac{\pi}{4}\frac{1}{N^2}cos^4(\alpha)
    \label{eq:lensradiometric}
\end{equation}
where L is the scene radiance, N is the lens aperture number and $\alpha$ is the angle between the principal ray and the ray through the pixel. The $cos^4(\alpha)$ term models natural vignetting, a process by which illumination decays towards the sensor edges. Additionally, some light is lost as it travels through the lens. This transmission loss depends on the quality and construction of the lens and usually ranges between $5\%$ and $20\%$~\cite{oelund2009}.

\subsection{Camera response}
Light that reaches the camera sensor is converted into an electrical signal. In our model, we assume the use of machine vision cameras with linear sensor response functions, though we note some consumer cameras have non-linear camera response functions, designed to mimic the chemical response of analog film. Grossberg et al.~\cite{grossberg_modeling_2004} studied the space of camera response functions. Debevec et al.~\cite{debevec_recovering_nodate} presented experimental methods to determine the camera response function from a set of images. Jiang et al.~\cite{jiang_what_2013} further modelled spectral sensitivity functions of color camera sensors and proposed experimental methods to obtain them from color board images. Our model assumes the sensor response is linearly dependent on the light intensity, with varying sensitivity to different wavelengths. The dependency of the sensor response on wavelength is described by the quantum efficiency curve. The total number of absorbed photons can be computed by dividing the spectrum energy, weighted with the quantum efficiency curve, by the energy of a photon:  
\begin{equation}
    \mu_{e} = \frac{At_{exp}}{\textit{h}c}\int_{\lambda_a}^{\lambda_b}\Phi(\lambda)\cdot\lambda\cdot \eta\left(\lambda\right) d\lambda 
    \label{eq:absorbedphotons}
\end{equation}
where A is the pixel area [$m^2$], $\Phi$ is the irradiance spectrum  [W/($m^2$nm)], $t_{exp}$ [s] is the exposure time, $h$ is Planck's constant, $c$ is the speed of light in air [m/s], $\lambda$ is the wavelength [m] and  $\eta\left(\lambda\right)$ is the sensor quantum efficiency as a function of wavelength. 
Following the EMVA1288 standard~\cite{jahne2010emva}, the digital sensor response signal $\mu_{y}$ can be computed as:
\begin{figure}[]
    \centering
    \includegraphics[width=0.9\columnwidth]{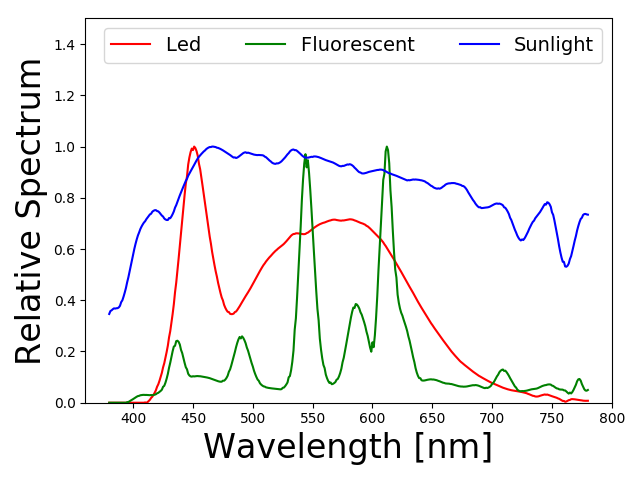}
    \caption{Radiance spectrum for different light types}
    \label{fig:lightsspectrum}
\end{figure}
\begin{equation}
    \mu_{y} = \mu_{y.dark} + K \mu_{e}
    \label{eq:adcconversion}
\end{equation}
where $\mu_{y.dark}$ is the sensor mean dark signal, and K is the system gain.  

The physical parameters for each sensor are published by camera manufacturers (eg.~\cite{flirblackflyspecs}) or can be obtained experimentally.

\subsection{Gain and Signal to Noise Ratio}

Similar to changing the ISO for film cameras, digital machine vision cameras can have a gain applied to the sensor response signal. This decreases the amount of scene light necessary to expose the image. However, the image noise is also amplified when a gain is applied, resulting in a reduction of the image \ac{SNR}. \ac{SNR} is an important consideration, especially for image tasks requiring feature matching~\cite{leclercq2003robustness}, and should be a parameter decided by the camera system designer. There are three sources of image noise: dark current noise, described by the normally distributed variance $\sigma_d^2$; quantization noise from the analog digital conversion, described by the normally distributed variance $\sigma_q^2$ and the overall system gain $K$; and shot noise inherent to light, described by the number of absorbed photons in the sensor $\mu_p$ and the sensor quantum efficiency $\eta$. The image \ac{SNR} is calculated as

\begin{equation}
SNR = \frac{\eta\mu_p}{\sqrt{\sigma_d^2+\sigma_q^2/K+\eta\mu_p}}.
\end{equation}

The camera system design tool allows setting a gain value and will display the calculated image \ac{SNR} for the target average exposure value.

\subsection{Operational Considerations}
Besides the physical characteristics of the water and selected equipment (camera, lens and lights), the operational requirements also highly influence the design space. The most significant of these requirements include: 
\begin{enumerate}[wide, labelwidth=!, labelindent=0pt]
    \item Minimum overlap between images: Overlap between consecutive images is required in order to perform photomosaics, 3D reconstructions or visual navigation. The amount of required overlap, together with the vehicle speed and working distance will determine the image acquisition frequency $f$:  
    \begin{equation}
        f = \frac{\textnormal{v}}{FOV_{x/y}(1-OVR)}
    \end{equation}
    where v is vehicle speed [m/s], $FOV_{x/y}$ is the spacial field of view of the image in the direction of motion [m], and OVR is the fraction of consecutive image overlap. 
    \item Focal \ac{DoF}: When running \ac{AUV} imaging surveys over rocky bottoms or coral reefs, it is frequent for the terrain height to vary significantly. It is desirable that the entire image remains in focus, so the required focal \ac{DoF} must be selected accordingly. Whether a pixel is in focus or not is determined by the circle of confusion, which describes the area of the sensor across which a point source of light is spread. Light rays originating within the focal range will project a circle of confusion on the sensor under an acceptable area threshold. The \ac{DoF} is controlled by an inverse relationship with the camera aperture. However, there is a trade off, as decreasing the size of the camera aperture decreases the amount of light that reaches the lens and therefore increases the required amount of light in the scene. The \ac{DoF} can be computed as:
    \begin{equation}
        DoF = \frac{2Ncf^{2}s^{2}}{f^{4}-N^{2}c^{2}s^{2}}
    \end{equation}
    where N is the lens aperture number, c is the diameter of the circle of confusion, f is the focal length, and s is the distance at which the camera is focused. 
    \item Motion blur: Motion blur is a great concern for underwater imaging platforms operating in low light. The amount of blur is dependent on the speed of the vehicle v [m/s], the camera field of view in the direction of motion $FOV_{x/y}$, the sensor resolution in the direction of motion $RES_{x/y}$, and the exposure time. The maximum exposure time $t_{exp}$ [s] to keep motion blur less than a set number of pixels $PIX_{blur}$ is given as:
    \begin{equation}
        t_{exp} = \frac{PIX_{Blur}\cdot FOV_{x/y}}{v \cdot RES_{x/y}}
    \end{equation}
    \item Spacial \ac{FOV}: The camera spacial \ac{FOV} or area covered by the image is influenced by lens selection and distance to the target D [m]. It can be computed as: \begin{equation}
        FOV_{x/y} = D*\frac{SS_{x/y}}{f} 
    \end{equation}
    where f is the lens focal length [mm], and $SS_{x,y}$ is the physical dimension of the sensor in x or y [mm].
\end{enumerate}

\section{Software}
\label{sec:software}
\begin{figure}
\centering
\includegraphics[width=\columnwidth]{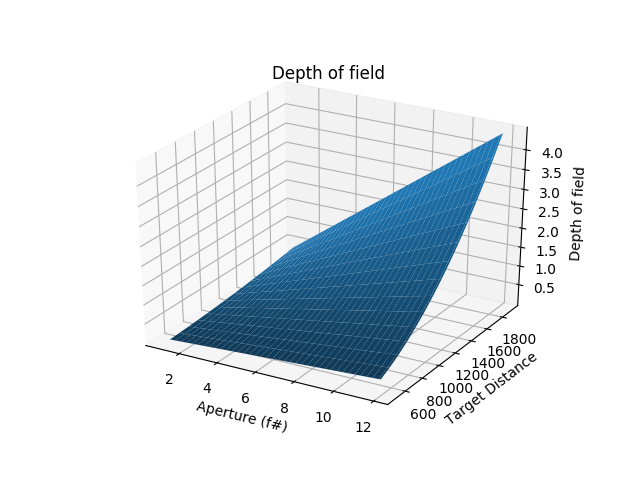}
\caption{Depth of field as a function of focus distance and aperture}
\label{fig:dof}
\end{figure}{}

\begin{figure}
\centering
\includegraphics[width=\columnwidth]{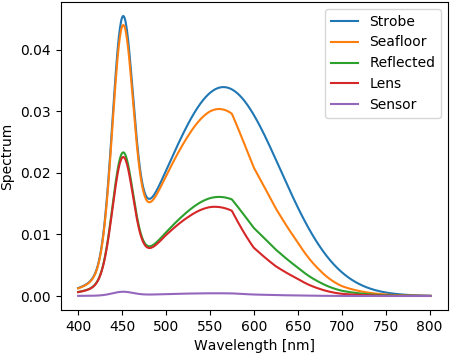}
\caption{Spectrum of light as it propagates through the water, attenuates, reflects and travels through the lens onto the sensor.}
\label{fig:software-spectrums}
\end{figure}{}

Taking the previously defined relations between sensors, lenses, light sources and water light propagation into account, users and designers of underwater camera systems may wish to answer questions like what sensor is best for a given operational profile? What are the lighting requirements for a specific camera? Or what aperture and shutter speed should be used for a given deployment scenario? In order to quickly answer questions like these we have developed an open source software design tool that performs parametric analysis of an underwater camera system. 

The tool allows the user to either input the light type and lumen intensity or load a custom light spectrum if available. Three Jerlov oceanic water types and five coastal water profiles are provided to analyze different attenuation rates, with an option to also load custom attenuation profiles. Lenses are defined by their focal length and their transmission loss, which may be specified either as a constant or by loading a custom wavelength dependent attenuation profile. Profiles are included with the program for five different camera sensors, and new sensors can easily be added if EMVA specifications are available from the manufacturer. The operational requirements of the camera system are specified in terms of the maximum acceptable motion blur, the minimum acceptable \ac{DoF}, the expected vehicle altitude and speed above the seafloor, and the desired percentage overlap of consecutive images. Other selectable parameters include the camera orientation with respect to the direction of vehicle motion, and the geometry of the camera housing viewport. With a given set of these parameters, the software computes the average camera response, minimum operational framerate, minimum exposure time, and minimum aperture number. In addition to the average camera response, the software can also generate visualizations of the parameter space for the given configuration. Figure~\ref{fig:dof} shows an example plot over a set of parameters, where the dependence of the \ac{DoF} on aperture and the distance to the imaged target is visualized.  Figure~\ref{fig:software-spectrums} shows an example plot of how the light spectrum is decayed as it propagates from the light source to the camera, helping contextualize the main sources of light reduction for a specified water environment. Similar plots may be generated by the software for the camera frame rate, exposure time or water attenuation profiles.




\section{Validation Experiments}
\label{sec:experiments}
\begin{figure}
    \centering
    \includegraphics[width=0.7\columnwidth]{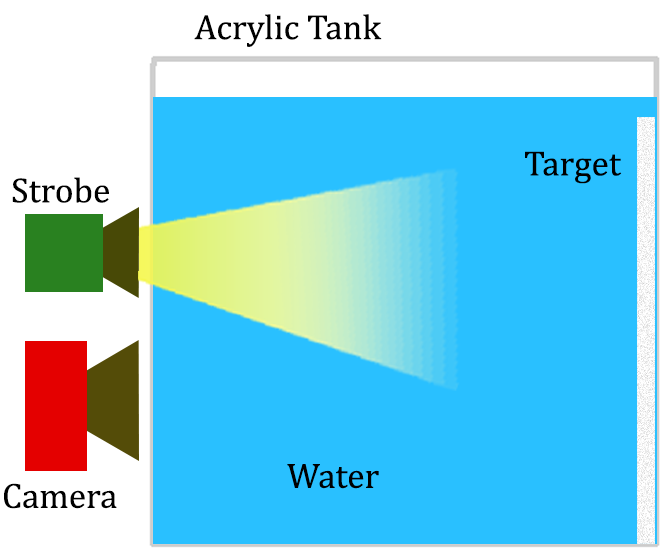}
    \caption{Experimental setup for verifying image formation model.}
    \label{fig:tank_setup}
\end{figure}

    

\begin{figure}
    \centering
    \includegraphics[width=\columnwidth]{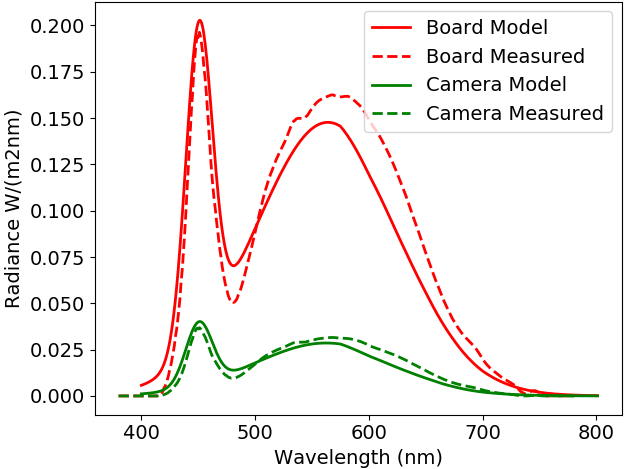}
    \caption{Comparison of measured and estimated light spectrums at both the target board as well as the camera position}
    \label{fig:exposure_experiments_spectrums}
\end{figure}

\begin{figure}
    \centering
    \includegraphics[width=\columnwidth]{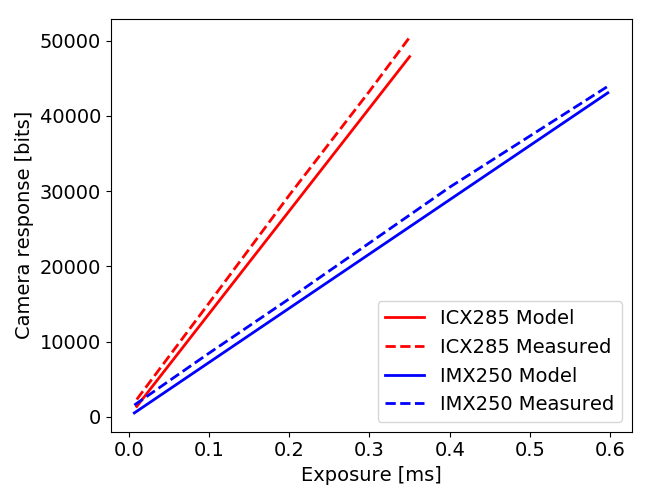}
    \caption{Measured and model predicted camera response curves for two different sensors under the same experimental conditions.}
    \label{fig:exposure_experiments_sensors}
\end{figure}%

\begin{figure}
    \centering
    \includegraphics[width=\columnwidth]{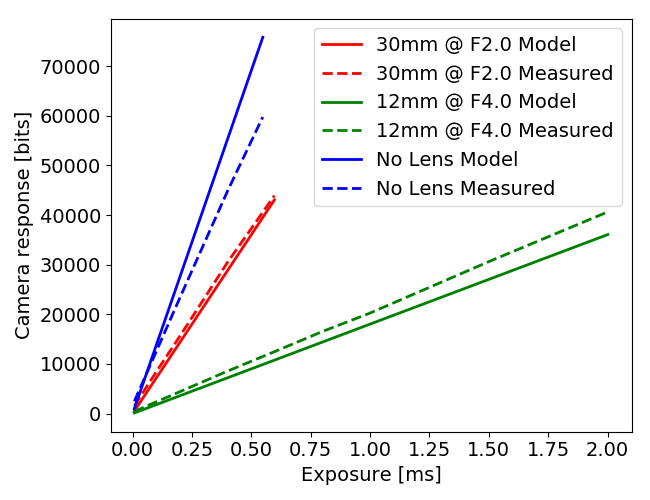}
    \caption{Camera response for two different lenses and without a lens.}
    \label{fig:exposure_experiments_lenses}
\end{figure}%

The camera response simulation pipeline is validated experimentally in a lab environment. We tested with two monochrome cameras, a Blackfly BFS-U3-51S5M from FLIR with a Sony IMX250 sensor and a Prosilica GT-1380 from Allied Vision with a Sony ICX285 sensor. The cameras were mounted on the outside of an $\SI{46}{\centi\meter}$~x~$\SI{46}{\centi\meter}$~x~$\SI{46}{\centi\meter}$ freshwater tank, with the camera axial direction perpendicular to the clear acrylic tank wall. A diffuse white target board was placed on the opposite side of the tank. Figure~\ref{fig:tank_setup} illustrates the experimental setup. Measurements were taken in dark ambient light conditions, with scene light being provided by a FixNeo25000DX $\SI{25}{\kilo\lumen}$ diving light positioned above the camera and against the outside tank wall. The light spectrum incident on the camera sensor was measured using a Sekonic SpectroMaster C-7000 lightmeter. The spectrometer was placed in a waterproof enclosure to perform spectrum measurements inside the tank

Figure~\ref{fig:exposure_experiments_spectrums} shows the measured light spectra versus those predicted by the model for a generic LED light source. The spectra are plotted for the light that was incident on the target surface, in red, and the light reflected back to the camera lens, in green. The model source light spectrum was calculated with the nominal luminous intensity provided by the manufacturer and a half beam angle of $40$deg, accounting for the change in beam angle from the manufacturer stated value due to refraction. The predicted model spectra, both at the target surface and at the camera lens, are very similar to the measured spectra in shape and size. Figure~\ref{fig:exposure_experiments_sensors} shows the response of the two different cameras to the light spectrum shown in Figure~\ref{fig:exposure_experiments_spectrums}. Both cameras had the same \SI{30}{\milli\meter} lens mounted with the aperture set at F$2.0$. The predicted responses from the model closely follow the measured values. We also compared the response of one camera with different lens and aperture configurations, including no lens, a \SI{30}{\milli\meter} lens with aperture F$2.0$, and a \SI{12}{\milli\meter} lens with aperture F$4.0$. Figure~\ref{fig:exposure_experiments_lenses} shows the measured versus the model predicted average camera responses for this experiment. For all camera experiments, the predicted responses from the model closely follow the measured responses, demonstrating the model is a good approximation of the real system and will give reliable predictions over the design space. 





\section{Conclusion and Future Work}
\label{sec:conclusion}
 
In this paper we have shown how underwater optical systems can be coarsely simulated by a set of simple equations, and we have developed a user-friendly interface to guide the component and parameter selections of such systems. The presented tool will enable researchers and engineers tasked with the development of underwater camera systems to better understand the available design space, analyze trade-offs in light, sensor and lens selection, and guide early design choices. 

Future work will include extending the model to consider systems with varying and multiple light source configurations and the addition of program features to aid in the focusing of domed and flat viewport camera systems.

\renewcommand{\bibfont}{\normalfont\small}
\printbibliography
\end{document}